\documentclass[twocolumn,abs,prb,longbibliography]{revtex4-1}
\usepackage{graphicx}
\usepackage{dcolumn}
\usepackage{float}
\usepackage{amsmath}
\usepackage{bm}

\newcommand{\comment}[1]{}

\begin{document}

%\preprint{}

\title{Spectroscopic signatures of phonons in high pressure superconducting hydrides}

\author{J. P. Carbotte$^{1,2}$}
\author{E. J. Nicol$^{3}$}
\email{enicol@uoguelph.ca}
\author{T. Timusk$^{1,2}$}
\affiliation{$^1$Department of Physics and Astronomy, McMaster
University, Hamilton, Ontario L8S 4M1, Canada}
\affiliation{$^2$The Canadian Institute for Advanced Research, Toronto, ON M5G 1Z8, Canada}
\affiliation{$^3$Department of Physics, University of Guelph,
Guelph, Ontario N1G 2W1, Canada} 
\date{\today}

\begin{abstract}{
The discovery of superconductivity at 203K in SH$_3$ is an important step toward higher values of critical temperature $T_c$. Predictions based on state-of-the-art density functional theory for the electronic structure, including one preceding experimental confirmation, showed the mechanism to be the electron-phonon interaction. This was confirmed in optical spectroscopy measurements. In the range of photon energies between $\sim 450$ and 600 meV in SH$_3$, the reflectance in the superconducting state is below that in its normal state. This difference decreases as temperature approaches $T_c$. Decreasing absorption with increasing temperature is opposite to what is expected in ordinary metals. Such an anomalous behavior can be traced back to the energy dependence of the superconducting density of states which is highly peaked at the energy gap value $\Delta$ but decays back to the constant normal state value as energy is increased, on a scale of a few $\Delta$, or by increasing temperature towards $T=T_c$. The process of phonon-assisted optical absorption is encoded with a knowledge of the temperature dependence of $\Delta$, which is also the order parameter characteristic of the superconducting state. Should the energy of the phonon involved be very large, of order 200 meV or more, this process offers the possibility of observing the closing of the superconducting order parameter with temperature at correspondingly very large energies. The very recent experimental observation of a $T_c\simeq 250$ K in LaH$_{10}$ has further heightened interest in the hydrides. Here, we compare the relevant phonon structure seen in optics with related features in the real and imaginary part of the frequency dependent gap, the quasiparticle density of states, the reflectance, the absorption, and the optical scattering rate. The phonon structures all carry information on the $T_c$ value and the temperature dependence of the order parameter, and can be used to confirm that the mechanism involved in superconductivity is the electron-phonon interaction.
}
\end{abstract}

\maketitle
% body of paper here

\section{Introduction}

Drozdov {\it et al.}\cite{Drozdov:2015,Drozdov:2014} first reported superconductivity in the sulfur hydrides at a record 203 K under a pressure of 155 GPa. In the cuprates, the maximum critical temperature is $T_c=133$ K at ambient pressure\cite{Schilling:1993} which can be raised\cite{Gao:1994} to 164 K under quasihydrostatic pressure of 45 GPa. These materials are, however, unconventional in that their superconducting order parameter has $d$-wave symmetry\cite{ODonovan:1995a,ODonovan:1995b} and their condensation into Cooper pairs is driven by correlations.\cite{Carbotte:2011} This is in contrast to the conventional superconducting materials of BCS theory which have an $s$-wave order parameter, possibly with some anisotropy.\cite{Leung:1976} The driving mechanism is the electron-phonon interaction in this case.

For many years it was not possible to raise the value of $T_c$ in conventional superconductors beyond 23.2 K in Nb$_3$Ge which led to the idea\cite{Cohen:1972} that due to lattice instability as the electron-phonon interaction increased, the $T_c$ may be limited to a maximum of 30 K or so. This all changed in 2001 with the discovery\cite{Nagamatsu:2001} of superconductivity in MgB$_2$ with $T_c\approx 40$ K which was rapidly established\cite{Kortus:2001,An:2001,Liu:2001,Kong:2001,Choi:2002} to be a two-band system\cite{Nicol:2005} with strong coupling of the quasi-two-dimensional $\sigma$-band to optical B-B (boron-boron) bond stretching phonons at 600 cm$^{-1}$. This leads to a large superconducting energy gap on this band with a smaller one on the $\pi$-band.

The possibility of high temperature electron-phonon superconductivity in metallic hydrogen was discussed by Ashcroft.\cite{Ashcroft:2004} Very recently, many papers, mostly based on density functional theory, have determined the electronic band structure, lattice dynamics, and electron-phonon spectral density $\alpha^2F(\omega)$ in hydrides\cite{Li:2014,Duan:2014,Errea:2015,Bernstein:2015,Papaconstantopoulos:2015,Flores-Livas:2016,Nicol:2015,Errea:2016}. The superconducting phase in the experiments of Ref.~\onlinecite{Drozdov:2015} was expected to be SH$_3$ with Im-3m structure as was verified in experiment.\cite{Einaga:2016} 

Recovery of detailed information on the electron-phonon spectral density from tunneling data has a long history which started with McMillan and Rowell\cite{McMillan:1965} for the specific case of Pb. This has allowed us to understand in great detail the properties of the superconducting state in many conventional materials\cite{Carbotte:1990} which deviate significantly from simple BCS expectations. It has also been possible to retrieve the same information on $\alpha^2F(\omega)$ in the case of Pb using optical techniques.\cite{Joyce:1970,Farnworth:1976} Optics has been particularly useful for the cuprates\cite{Basov:2005} and has also been applied to other cases\cite{Marsiglio:1998,Mori:2008}, including MgB$_2$ (Ref.~\onlinecite{Hwang:2014}). Very recently, Capitani {\it et al.}\cite{Capitani:2017} measured the reflectance of SH$_3$ at 155 GPa in a diamond anvil cell (DAC) and so probed the electron-phonon interaction in the system establishing a new high energy scale for its superconductivity. These experiments leave no doubt that we are dealing with conventional electron-phonon superconductivity.\cite{Marsiglio:2008}

Recent important new developments include, among others, work\cite{Liu:2017} on 
LaH$_{10}$, a hydrogen-rich superhydride with clathrate-type structure\cite{Somayazulu:2019} in which the H-H distance is $\sim 1.1$\AA\ at a pressure of $\sim 210$ GPa with an expected $T_c$ value of order 220 K for La-H and 300 K for Y-H. Hydrogen-rich crystals in this series have already been synthesized in at least two laboratories.\cite{Somayazulu:2019,Drozdov:2018a,Drozdov:2019} The samples have somewhat different values of $T_c$ which also vary with pressure. In Ref.~\onlinecite{Drozdov:2019}, the crystal structure is Fm-3m at a pressure of $\sim 170$ GPa with $T_c=250$ K, as determined through measurements of zero resistance, isotope shift, and reduction in $T_c$ on application of an external magnetic field.

In this paper, we describe how optical conductivity data, taken as a function of photon energy $\Omega$ at various temperatures up to $T_c$ and beyond, can be employed to derive information on the electron-phonon spectral density $\alpha^2F(\omega)$ versus phonon energy. Within Migdal-Eliashberg theory,\cite{Carbotte:1990,Marsiglio:2008}
this function has built into it all the information on the electrons, phonons and electron-phonon coupling strength needed to fully characterise the mechanism which drives the condensation into Cooper pairs. In particular, we emphasize how the phonon structure becomes embedded in the superconducting gap function itself, which acquires nontrivial energy dependence of its own and has both real and imaginary parts. The complex gap function $\Delta(T,\omega)$ follows from the solution of the Eliashberg equations which are a set of two nonlinear coupled integral equations.\cite{Carbotte:1990,Marsiglio:2008} The gap determines spectroscopic properties and consequently phonon structure appears in such properties and, as we will argue, provides a mechanism by which the energy and temperature variation of the gap, which is on the scale of 50 meV in the high $T_c$ hydrides, can be seen at much higher energies of order $2\Delta$ plus the maximum phonon energy $\omega_{\rm max}$.

Eliashberg theory is well documented in the literature to which the reader is referred.\cite{Carbotte:1990,Marsiglio:2008} The details of the theory will not be repeated here. In section~II, we present our solutions of the Eliashberg equations for the  specific case of SH$_3$. The solutions of these equations involve the gap renormalization function $\Delta(T,\omega)$ and the renormalization function $Z(T,\omega)$, both of which are complex functions. The electron-phonon spectral density $\alpha^2F(\omega)$, which is input to these equations, is taken to be the one that was used before in our previous work\cite{Nicol:2015,Capitani:2017} on SH$_3$ at 200 GPa and is based on the anharmonic density functional results of Errea {\it et al.}\cite{Errea:2015}. We compare phonon structures which appear in both the real and imaginary parts of the output functions $\Delta(T,\omega)$ and $Z(T,\omega)$ with particular emphasis on the energy scale. In section~III, we consider the temperature dependencies for the  energy gap $\Delta_0(T)$ obtained from the solution of the equation
$\Delta_0={\rm Re}\Delta(T,\omega=\Delta_0)$ and find that it deviates somewhat from the classic BCS mean field result. This also holds when the normalised inverse squared penetration depth $[\lambda(0)/\lambda(T)]^2$ is considered. These two temperature variations are compared with dependencies of both the real and imaginary
parts of $\Delta(T,\omega)$ taken at some fixed value of $\omega$. Section IV is devoted to a discussion of the quasiparticle density of states in the superconducting state $N(T,\omega)/N(0)$, with $N(0)$ the normal state value at the Fermi energy. Section~V involves the optics and we show results for the reflectance $R(T,\omega)$, the absorption $A(T,\omega)$, and the optical scattering rate $1/\tau_{\rm op}(T,\omega)$. We provide a summary and conclusion in section~VI.

\section{The gap and renormalization functions}

\begin{figure}
\includegraphics[width=0.9\linewidth]{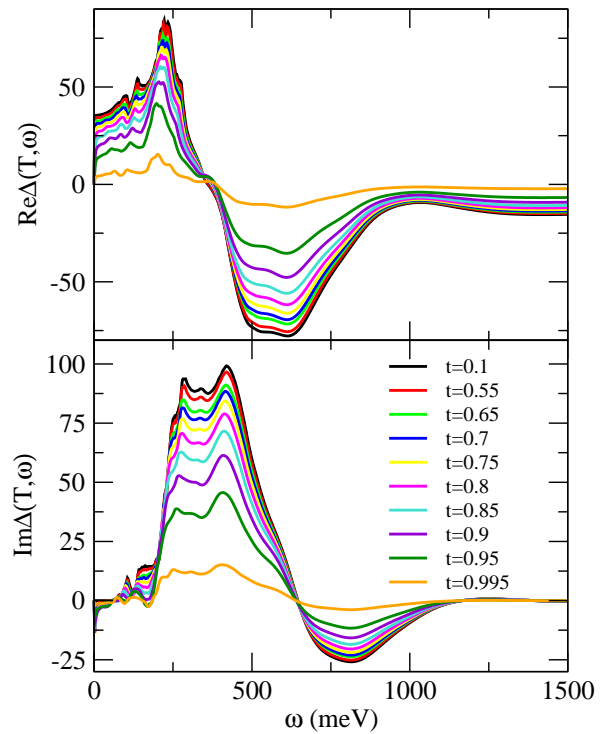}
\caption{
The real and imaginary part of the Eliashberg gap function $\Delta(T,\omega)$
as a function of energy $\omega$, top and bottom frames, respectively. 
Shown are curves for 10 temperatures up to $t=T/T_c\sim 1$ as labeled in the figure. These curves are dominated in their $\omega$ dependence by the phonon structures coming from the electron-phonon spectral density $\alpha^2F(\omega)$. At fixed frequency, their temperature dependence is encoded with information on the mean field order parameter.
}\label{fig1}
\end{figure}

\begin{figure}
\includegraphics[width=0.9\linewidth]{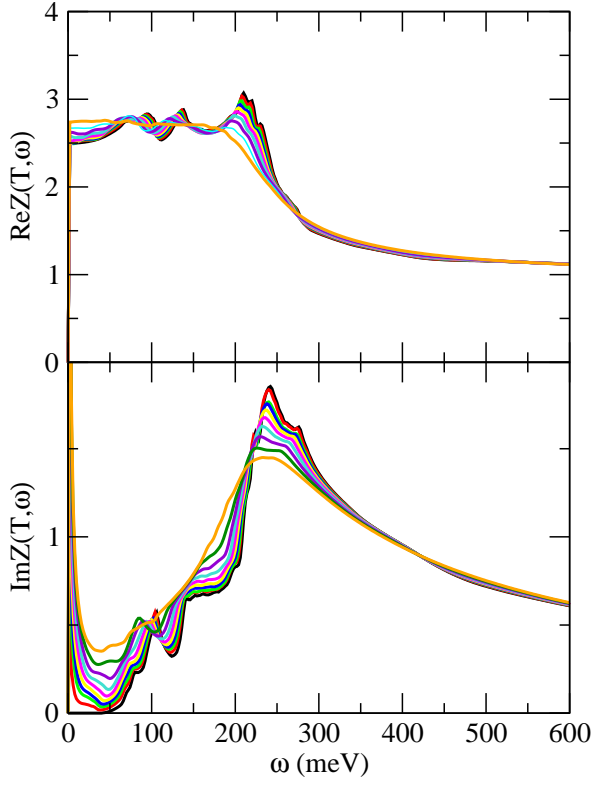}
\caption{
Same as for Fig.~\ref{fig1} but now the real and imaginary parts of the renormalization function $Z(T,\omega)$ are shown.
}\label{fig2}
\end{figure}

The Eliashberg equations written on the real\cite{Marsiglio:2008,Carbotte:1990} frequency axis deal with a frequency and temperature dependent gap\cite{Marsiglio:1988} $\Delta(T,\omega)$ and renormalization $Z(T,\omega)$. Both are complex functions. The inputs to these equations, which characterise the material involved, are the electron-phonon spectral function $\alpha^2F(\omega)$ of density functional theory\cite{Errea:2015} and a Coulomb repulsion denoted by $\mu^*$, here taken to be 0.18.\cite{Nicol:2015} In our work here, we begin by solving the Eliashberg equations on the imaginary frequency axis and then analytically continue to the real frequency axis. The Eliashberg equations on the imaginary frequency axis are given as:\cite{Carbotte:1990}
\begin{equation}
\tilde\omega_n = \omega _n+\pi T
 \sum_{m}\,\lambda(n-m)
{\tilde\omega_m
\over\sqrt{\tilde\omega_m^2+
\tilde\Delta _m^2}}
\quad 
\end{equation}
and
\begin{equation}
\tilde\Delta _n= \pi T
 \sum_{m}\,[\lambda(n-m)-\mu^*\theta(\omega_c-\mid\omega_m\mid)]
{\tilde\Delta _m
\over\sqrt{\tilde\omega_m^2+
\tilde\Delta _m^2}},
\end{equation}
with
\begin{equation}
\lambda(n-m)=\int _0^\infty\,{2\Omega\,\alpha ^2F(
\Omega)\over \Omega ^2+(\omega _n-\omega _m)^2}\, d\Omega ,
\end{equation}
where
$\tilde\omega_n\equiv\tilde\omega(i\omega_n)\equiv Z(i\omega_n)\omega_n$ and $\tilde\Delta_n\equiv\tilde\Delta(i\omega_n)\equiv Z(i\omega_n)\Delta(i\omega_n)$. Here,
$\omega_n=\pi T(2n-1)$ are the fermionic Matsubara frequencies for $n=0, \pm 1, \pm 2, ...$ and $\omega_c$ is a high frequency cutoff typically taken as six times the maximum phonon frequency in the $\alpha^2F(\omega)$ spectrum.
For a particular $\alpha^2F(\Omega)$ and $\mu^*$, these equations are iterated to convergence for a chosen temperature and then the solutions for $\tilde\omega_n$ and $\tilde\Delta_n$ are analytically continued to the real axis forms via iterating the following equations:\cite{Marsiglio:1988}
\begin{widetext}
\begin{eqnarray}
  \displaystyle
  \tilde\omega(\omega)&=  \displaystyle
\omega+i\pi T\sum_{m=0}^\infty
[\lambda(\omega-i\omega_m)-\lambda(\omega+i\omega_m)]
{\tilde\omega(i\omega_m)\over 
\sqrt{\tilde\omega^2(i\omega_m)+\tilde\Delta^2(i\omega_m)}}\cr
&+   \displaystyle
i\pi\int_{-
\infty}^{+\infty} dz
\alpha^2F(z)[n(z)+f(z-\omega)]
{\tilde\omega(\omega-
z)\over\sqrt{\tilde\omega^2(\omega-z)-\tilde\Delta^2(\omega-z)}}
\end{eqnarray}
and
\begin{eqnarray}
  \displaystyle
  \tilde\Delta(\omega)&=  \displaystyle
i\pi T\sum_{m=0}^{\infty}
[\lambda(\omega-i\omega_m)+\lambda(\omega+i\omega_m)-2\mu^*]
{\tilde\Delta(i\omega_m)\over\sqrt{\tilde\omega^2(i\omega_m)+
\tilde\Delta^2(i\omega_m)}}\cr
&+\   \displaystyle
i\pi\int_{-\infty}^{+\infty}dz
\alpha^2F(z)[n(z)+f(z-\omega)]{\tilde\Delta(\omega-
z)\over \sqrt{\tilde\omega^2(\omega-z)-\tilde\Delta^2(\omega-z)}},
\end{eqnarray}
\end{widetext}
with
\begin{equation}
\lambda(\omega)=-\int_{-\infty}^{+\infty}{d\Omega\alpha^2F(\Omega)\over \omega-
\Omega+i0^+}\quad ,
\end{equation}
where
$n(\omega)=1/[{\rm exp}(\beta\omega)-1]$ and 
$f(\omega)=1/[{\rm exp}(\beta\omega)+1]$, are the Bose-Einstein and Fermi-Dirac distributions, respectively, with $\beta=1/(k_BT)$ 
and $k_B$, the Boltzmann constant. Here, $\tilde\Delta(\omega)\equiv\Delta(T,\omega)$ and $\tilde\omega(\omega)\equiv\omega Z(T,\omega)$. To access the BCS limit numerically through these equations, one can take $\alpha^2F(\Omega)$ to be a delta function positioned at very high frequency or, as aluminum is a classic BCS superconductor, one may use the aluminum $\alpha^2F(\Omega)$ spectrum\cite{Carbotte:1990}. In the BCS limit $Z(T,\omega)$
is replaced by the normal state renomalization function $Z_N$. For the electron-phonon problem, this would be the value $1+\lambda$, where $\lambda$ is the electron-phonon mass renormalization parameter. If the normal state renormalizations are ignored, then $Z(T,\omega)=1$.
Also, in the BCS limit, ${\rm Re}\Delta(T,\omega)=\Delta_0(T)$ (a constant at each temperature) up to a cutoff which is the Debye energy, and zero thereafter. ${\rm Im}\Delta(T,\omega)$ is zero in this limit. One can take the imaginary axis equations above and apply a ``two-square-well'' model, which leads to a renormalized BCS form for the BCS gap and $T_c$ equations.\cite{Carbotte:1990,Nicol:2005} Further discussion about reducing the real-axis equations to BCS can be found in Ref.~\onlinecite{Carbotte:1990}.

In Fig.~\ref{fig1}, top frame, we show results for ${\rm Re}\Delta(T,\omega)$ as a function of $\omega$ for various values of reduced temperature $t\equiv T/T_c$  (as labeled in the figure) up to $t\sim 1$ (or $T\sim T_c$). The lower frame gives the corresponding results for ${\rm Im}\Delta(T,\omega)$. In Fig.~\ref{fig2}, we present equivalent results for the real and imaginary parts of the renormalization $Z(T,\omega)$. First, as mentioned above, for a BCS superconductor the real part of the gap would have no $\omega$ dependence, except for a cutoff at the Debye energy $\omega_D$, and its imaginary part would be zero. This is in sharp contrast to what we have in SH$_3$. The large amount of structure is due to the electron-phonon spectral density $\alpha^2F(\omega)$ and reflects its $\omega$ dependence. In SH$_3$, the maximum phonon energy is about 250 meV while the gap edge at zero temperature is 36.5 meV. This is defined by the solution of the equation $\Delta_0(T)={\rm Re}\Delta[T,\omega=\Delta_0(T)]$. Note that there are two separate regions of variation for the phonon-induced structure in ${\rm Re}\Delta(T,\omega)$. Below about 370 meV, the boson structure is largest (positive) at the lowest temperature while above it is most negative at $t=0.1$. Also note that the variations of these structures with temperature are most rapid as $T$ approaches $T_c$, which is characteristic of a second order phase transition. In the lower frame of 
Fig.~\ref{fig1} we see very much the same behavior for ${\rm Im}\Delta(T,\omega)$ although the crossing of the curves from one region to the other is at a different energy ($\sim 625$ meV). Finally, we note that the renormalization function $Z(T,\omega)$ shown in Fig.~\ref{fig2} also shows considerable phonon structure but this is confined to a lower energy range. As mentioned earlier, in unrenormalized BCS, ${\rm Re}Z(T,\omega)=1$ and ${\rm Im} Z(T,\omega)=0$. In Eliashberg theory, we see at very high frequency ${\rm Re}Z(T,\omega)$
approaches the unrenormalized value of 1 and at $\omega\to 0$ and $T\to 0$, 
${\rm Re}Z(T,\omega)\to 1+\lambda$.

\section{Temperature dependences}

\begin{figure}
\includegraphics[width=0.9\linewidth]{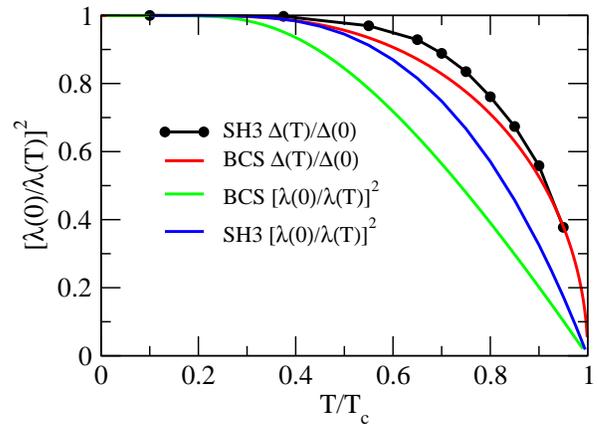}
\caption{
The temperature dependence of the normalised BCS energy gap $\Delta(T)$ (red curve) compared with the SH$_3$ gap edge (black dots with connecting curve). Also shown are results for the corresponding normalised inverse square of the penetration depth $\lambda(T)$, with the BCS case in green and the SH$_3$ result in blue. 
}\label{fig3}
\end{figure}

Next we concentrate on temperature dependences involved in the gap variations. In BCS, the gap edge $\Delta(T)$ carries information on the second order phase transition that is involved in the Cooper pair condensation. In such a case, the order parameter (or the gap) will be flat, almost temperature independent, at low $T$ but then decrease rapidly as $T$ approaches $T_c$. This is shown in Fig.~\ref{fig3}. The black curve with the dots is the result of our Eliashberg equations for the gap edge compared with the BCS result (red curve). We see that these variations, while not quite the same in magnitude, nevertheless track each other well. Indeed, the slight enhancement of the SH$_3$ gap is typical of strong electron-phonon coupling superconductors, a class to which SH$_3$ belongs. Also, shown in the same figure are additional results for the London penertration depth $\lambda(T)$, plotted in the form of the superfluid density $[\lambda(0)/\lambda(T)]^2$. This quantity is evaluated from:\cite{Carbotte:1990}
\begin{equation}
\biggl[\frac{\lambda(0)}{\lambda(T)}\biggr]^2=\frac{T}{2}\sum_{n=1}^{\infty}
\frac{\Delta^2(i\omega_n)}{Z(i\omega_n)[\omega_n^2+\Delta^2(i\omega_n)]^{3/2}},
\label{eq:pendepth}
\end{equation}
where 
$1/\lambda^2(0)=\omega_p^2=4\pi ne^2/m$, with $e$ and $m$ the electron charge and effective mass, respectively, and $n$ the electron density.
The blue curve is our calculation for SH$_3$ while the green curve is the classic BCS result. Again, SH$_3$ deviates in a manner consistent with strong coupling effects. What is important to emphasize here is that while $\Delta(T)/\Delta(0)$ and $[\lambda(0)/\lambda(T)]^2$ do not have exactly the same $T$ dependence, both could be taken to be the underlying order parameter associated with the superconducting state.

\begin{figure}
\includegraphics[width=0.9\linewidth]{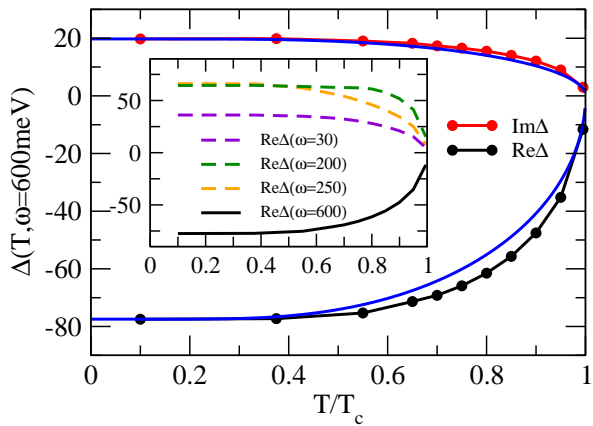}
\caption{
The temperature variation for the Eliashberg gap function obtained from numerical data in Fig.~\ref{fig1} at a fixed energy $\omega=600$ meV. Shown are ${\rm Re}\Delta(T,\omega=600$meV) (black dots connected by the black curve) and ${\rm Im}\Delta(T,\omega=600$meV) (red dots with connecting red curve). The blue curves are scaled BCS temperature variations for comparison. The inset shows further numerical results for only Re$\Delta(T,\omega)$ at fixed  energy $\omega =$ 600 meV (black) (repeated from the main frame), 250 meV (orange dashed), 200 meV (dark green dashed) and 30 meV (violet dashed).
}\label{fig4}
\end{figure}

Temperature dependences that give information on the superconducting state are not confined to the zero energy limit but are also embedded in the gap at finite frequencies. This is shown in Fig.~\ref{fig4} where we show our results for 
${\rm Re}\Delta(T,\omega)$ 
and ${\rm Im}\Delta(T,\omega)$ at $\omega=600$ meV (black with dots and red with dots, respectively) and compare with BCS temperature variations that have been scaled to approximately match the curves (blue curves). These results make it clear that we can see the order parameter involved at large energies way beyond the scale of the energy gap $\sim 36$ meV. The inset of Fig.~\ref{fig4} presents additional results 
for the case of ${\rm Re}\Delta(T,\omega)$ 
for different choices of fixed $\omega$, namely $\omega=600$ meV (repeated from the main frame), 250 meV, 200 meV, and 30 meV.

\section{Quasiparticle density of states}

\begin{figure}
\includegraphics[width=0.9\linewidth]{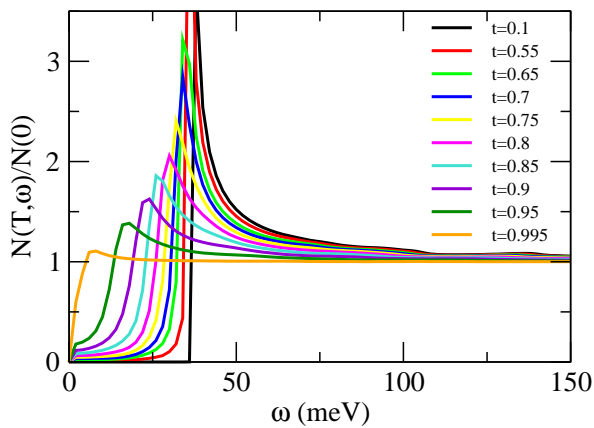}
\caption{
The quasiparticle density of states $N(T,\omega)$ as a function of energy $\omega$ for the ten temperatures 
labeled in the figure, ranging from $t=T/T_c=0.1$ to $t=0.995$. The low energy region around the gap is emphasized. In SH$_3$, the gap at low $t$ is 36.5 meV. Note the small (in comparison) structures at higher energies $\sim 100$ meV or so. These will be emphasized in Fig.~\ref{fig6}.
}\label{fig5}
\end{figure}

\begin{figure}
\includegraphics[width=0.9\linewidth]{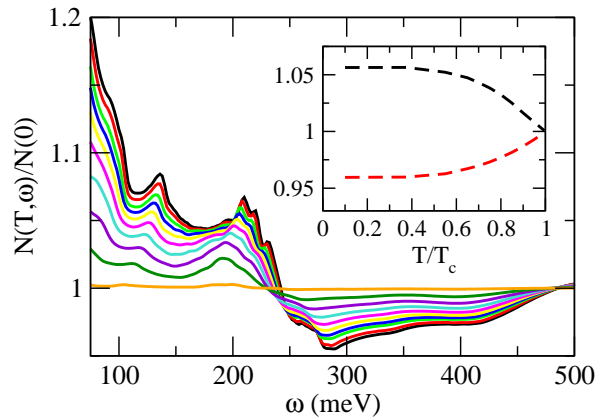}
\caption{
The quasiparticle density of states $N(T,\omega)$ as a function of energy $\omega$ for the same temperatures as in Fig.~\ref{fig5}. Here, the structure in this quantity up to 300 meV is emphasized. In the  inset, the temperature variation of $N(T,\omega)$ is shown for two values of fixed $\omega$: $\omega =$ 150 meV (black dashed) and 300 meV (red dashed). Both are equal to 1 at $T=T_c$ (normal state) as the superconducting order parameter has closed.
}\label{fig6}
\end{figure}

The superconducting quasiparticle density of states $N(T,\omega)$ is given by\cite{McMillan:1965}
\begin{equation}
\displaystyle {N(T,\omega)\over N(0)}={\rm Re}\biggl\{{\omega\over \sqrt{\omega^2-\Delta(T,\omega)^2}}\biggr\},
\label{eq:dos}
\end{equation}
where $N(0)$ is the density of states at the Fermi energy in the normal state. It has played a central role in our knowledge of the electron-phonon spectral density in conventional superconductors.\cite{Marsiglio:2008,McMillan:1965,Carbotte:1990} In Fig.~\ref{fig5}, we show the results for SH$_3$ for $N(T,\omega)/N(0)$ as a function of energy $\omega$ at the ten different temperatures labeled in the figure. The emphasis is on the low energy region around the gap edge. At low $T$ ($t=T/T_c=0.1$, black curve), we see a prominent peak in the density of states at the energy gap value (basically zero temperature). As temperature is increased towards $T_c$, the peak smears and is shifted to lower values. The curves clearly contain information on the temperature variation of the underlying order parameter on the energy scale of a few $\Delta$ in the range $T\le T_c$. Similar information, however, can also be found when one looks at higher energy, and this has not been emphasized in the past. In Fig.~\ref{fig6}, we show the region of $N(T,\omega)$ beyond $\omega\sim 100$ meV on a scale that emphasizes the phonon structure. Below $\omega\approx 250$ meV, the boson structure decreases as $T$ increases and $N(T,\omega)/N(0)$ tends towards 1 from above, while in the region above 250 meV, it tends towards 1 from below. The inset gives values of $N(T,\omega)/N(0)$ versus $T$ at fixed value of $\omega$, namely, $\omega=$ 150 meV (black dashed curve) and 300 meV (red dashed curve). It is clear that information on the second order phase transition is also encoded in the density of states at high $\omega$ values much larger than the energy gap scale, and measurements in this range would provide an alternate route to this information.

\section{Optical conductivity}

\begin{figure}
\includegraphics[width=0.9\linewidth]{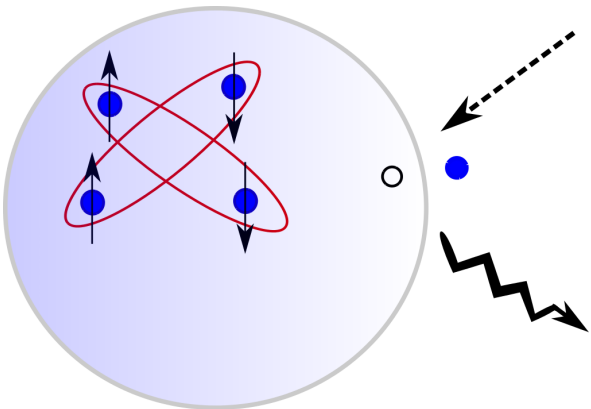}
\caption{
Schematic diagram of a phonon-assisted absorption process. A photon  (dashed arrow) creates a hole-particle pair out of the Cooper pair condensate while at the same time a phonon of energy $\omega_0$ is created (wiggled arrow). This process could be one where, for example, in breaking the Cooper pair the photon excites a quasiparticle to high energy which decays by emitting a phonon. This allows the energy of the photon involved to be at large energy of order twice the energy gap plus $\omega_0$ so that such processes carry information on both the condensate and on the energy scale associated with the phonon spectral density. Consequently, superconductivity is seen on an energy scale much larger than the gap.
}\label{fig7}
\end{figure}

Next we consider optical properties. The dynamic longitudinal conductivity $\sigma(T,\Omega)$ as a function of photon energy $\Omega$ and temperature $T$ is given by\cite{Marsiglio:2008,Akis:1991,Carbotte:2018}
\begin{align}
\sigma(T,\Omega)&={i\omega^2_p\over 4\pi\Omega}\biggl\{ \int_{0}^{\infty} 
d\omega {\rm tanh}\biggl({\omega\over 2k_BT}\biggr) 
[J_1(\omega,\Omega)+J_2(\omega,\Omega)]\nonumber\\
&+\int_{-\Omega}^Dd\omega {\rm tanh}\biggl({\omega+\Omega\over 2k_BT}\biggr) 
[J^*_1(\omega,\Omega)-J_2(\omega,\Omega)]\biggr\},\label{eq:sigma}
\end{align}
with $D$ a large cutoff taken to infinity for large electronic bandwidth. $\omega_p$ is the plasma frequency and 
\begin{align}
2J_1(\omega,\Omega)=& {1-M(\omega)M(\omega+\Omega)-P(\omega)P(\omega+\Omega)\over
E(\omega)+E(\omega+\Omega)+i/\tau_{\rm imp}},\label{eq:sigmaJ1}\\
2J_2(\omega,\Omega)=& {1+M^*(\omega)M(\omega+\Omega)+P^*(\omega)P(\omega+\Omega)\over
E^*(\omega)-E(\omega+\Omega)-i/\tau_{\rm imp}},
\label{eq:sigmaJ2}
\end{align}
where $*$ indicates the complex conjugate. 
The static impurity scattering $1/\tau_{\rm imp}$ enters through the denominators of Eqns.~(\ref{eq:sigmaJ1}) and (\ref{eq:sigmaJ2}) as shown. The various quantities are
\begin{equation}
E(\omega)=\sqrt{\tilde\omega^2(\omega)- \tilde\Delta^2(\omega)},
\end{equation}
with $E(-\omega)=-E^*(\omega)$, and
\begin{equation}
M(\omega)={\tilde\omega(\omega)\over E(\omega)}
\end{equation}
and 
\begin{equation}
P(\omega)={\tilde\Delta(\omega)\over E(\omega)},
\end{equation}
with $\tilde\Delta(\omega)=Z(\omega)\Delta(\omega)$ and $\tilde\omega(\omega)=\omega Z(\omega)$. We have suppressed the label of $T$ for the explicit dependence on temperature. 
Both the real and imaginary parts of the conductivity are given by
Equation~(\ref{eq:sigma}).

As we will examine here below,
the prominent phonon structures of the previous section also appear in the optical conductivity and are translated to related quantities: the reflectance, absorption and optical scattering rate.
To aid in understanding why this would occur, consider 
Fig.~\ref{fig7} which is illustrative and describes phonon-assisted optical absorption processes. It shows a photon of energy $\Omega$ impinging on a Cooper pair condensate with the result that two quasiparticles (an excited hole-particle pair) are created as well as a phonon of energy $\omega_0$. Even when the energy of the particle-hole pair is of order twice the energy gap, the energy of the photon involved can be very large compared with the scale of a few $\Delta$. As an example, this can occur because the photon could excite the quasiparticle to high energy where it decays back to the gap edge via phonon emission. The availability of phonons with large value of $\omega_0$ effectively boosts the energy scale of the photon involved in this process (a few $\Delta +\omega_0$) thereby allowing the energy and temperature variations typical of the superconducting state, a few $\Delta$ in magnitude, to be seen by probing optical properties at high energies of order 300 meV or larger. This is the physics we wish to stress in this paper. 

\begin{figure}
\includegraphics[width=0.9\linewidth]{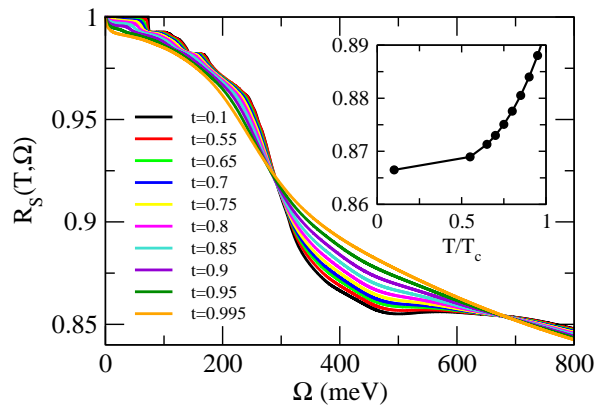}
\caption{
The superconducting state reflectance $R_s(T,\Omega)$ for SH$_3$ as a function of photon energy $\Omega$ shown for the temperatures $t=T/T_c$ labeled in the figure. The crossing of curves just below 300 meV is close to the energy of twice the gap at zero temperature plus the maximum phonon energy ($\omega_{\rm max}$) for the system. In the inset, $R_s(T,\Omega=400$ meV) is plotted as a function of temperature and this quantity shows mean field temperature dependence.
}\label{fig8}
\end{figure}

Results for the superconducting state reflectance $R_s(T,\Omega)$ of SH$_3$ as a function of photon energy $\Omega$ at several temperatures are given in Fig.~\ref{fig8}. We see that in this quantity there is a crossing of the curves around $\sim 300$ meV which also corresponds roughly to twice the gap value at $T=0$ plus the maximum phonon energy in the electron-phonon spectral density $\alpha^2F(\omega)$ of SH$_3$ in the anharmonic approximation, which is $\sim 250$ meV. We also see boson structures below the energy scale as well as above, where it extends to 700 meV or so. In the region 300 to 700 meV, the reflectance shows an increase with increasing temperature which is precisely the opposite of what is expected for a normal metal where increasing $T$ should increase the absorption and so decrease the reflectivity. Encoded in this temperature dependence of $R_s(T,\Omega)$ at some well chosen fixed value of $\Omega$, say $\Omega=400$ meV for SH$_3$, is the the mean field temperature dependence of superconducting properties. This is illustrated in the inset where we have plotted $R_s(T,\Omega=400$ meV) versus temperature. We see a flat behavior at small value of $T$ and a rapid growth towards value 1 around $T\lesssim T_c$. While this temperature variation is somewhat different in detail to that for the gap or for the penetration depth seen in Figs.~\ref{fig3} and \ref{fig4}, it nevertheless could serve as an order parameter and it is this variation characteristic of the superconducting state which is seen at high energies in the phonon-assisted optical spectroscopy.

\begin{figure}
\includegraphics[width=0.9\linewidth]{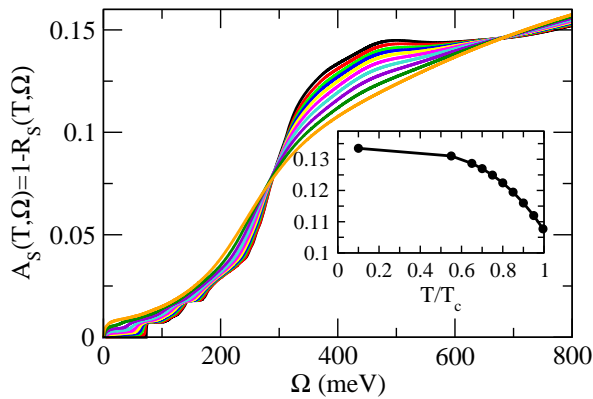}
\caption{
Same as for Fig.~\ref{fig8} but now the absorption $A_s(T,\Omega)=1-R_s(T,\Omega)$ is shown.
}\label{fig9}
\end{figure}

\begin{figure}
\includegraphics[width=0.9\linewidth]{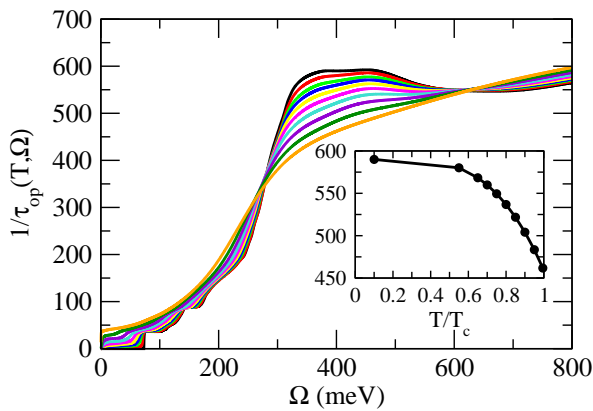}
\caption{
Same as for Fig.~\ref{fig8} but now the optical scattering rate $1/\tau_{\rm op}(T,\Omega)$ is plotted. These last three figures follow the same trends and all can be used to probe the superconducting state and reveal that it is electron-phonon-driven with a very high phonon energy scale.
}\label{fig10}
\end{figure}

Figs.~\ref{fig9} and \ref{fig10} illustrate very much the same thing. In Fig.~\ref{fig9} we show results for the absorption in the superconducting state 
defined by $1-R_s(T,\Omega)\equiv A_s(T,\Omega)$ 
as a function of $\Omega$ for the same set of temperatures as we showed in Fig.~\ref{fig8} and the inset gives $A_s(T,\Omega=400$ meV) as a function of $T$. It has its own temperature dependence but also shows a flat region at low $T$ and a much sharper dependence on $T/T_c$ as the phase transition from the superconducting to normal state is reached.

In Fig.~\ref{fig10}, we show additional results for the optical scattering rate extensively used in the literature to present superconducting state
optical data. In terms of the conductivity $\sigma(T,\Omega)$, the optical scattering rate is defined as\cite{Basov:2005,Marsiglio:1998,Mori:2008,Hwang:2014}
\begin{equation}\displaystyle
{1\over \tau_{\rm op}(T,\Omega)}={\omega_p^2\over 4\pi} {{\rm Re}\sigma(T,\Omega)\over
[{\rm Re}\sigma(T,\Omega)]^2+[{\rm Im}\sigma(T,\Omega)]^2},
\label{eq:tauop}
\end{equation}
where $\omega_p^2$ is the plasma frequency. In Fig.~\ref{fig10}, the results show that this quantity behaves very much as the absorption of Fig.~\ref{fig9} with respect to $\Omega$ dependence and temperature variation (see inset to Fig.~\ref{fig10} for the optical scattering rate at fixed $\Omega=400$ meV).

\section{Summary and conclusions}

We have presented a detailed account of how phonon structure in an Eliashberg superconductor manifests itself in various properties. To be specific, we considered SH$_3$, a recently discovered, very high $T_c$ superconductor. The information on the electron-phonon interaction enters through its spectral density $\alpha^2F(\omega)$ known from density functional theory (DFT).  It is encoded in to the gap function $\Delta(T,\omega)$ which is energy and temperature dependent. Similar structures are also present in the  mass renormalization function $Z(T,\omega)$. For the gap, such structures extend to very high energies much beyond the BCS gap energy scale and persist to $\omega$ of order several 100 meV. It is shown that for a well-chosen fixed value of $\omega$, the temperature  dependence of both the real and imaginary part of $\Delta(T,\omega)$ shows variations which reveal the mean field temperature dependence of a second order phase transition, as does the gap edge $\Delta(T)/\Delta(0)$ in BCS or alternatively the inverse square penetration depth $[\lambda(0)/\lambda(T)]^2$. For SH$_3$, these variations are slightly modified by strong coupling effects.

The boson structures seen in the gap function also are transferred to other properties such as the quasiparticle density of states $N(T,\omega)$ which can be used to establish that the superconductivity has its origin in a phonon mechanism with a very high value of maximum phonon energy $\sim 250$ meV. The optical properties show the same trends and we discussed in detail the reflectance, the absorption, and the optical scattering rate, with particular emphasis on phonon-assisted optical absorption in the range of photon energy $\Omega$ of order 600 meV. The temperature dependence  of the data in this range reveals the second order phase transition involved in the condensation into Cooper pairs.

\begin{acknowledgments}

This work has been supported by the Natural Sciences and Engineering Research Council of Canada (NSERC) and by the
Canadian Institute for Advanced Research.

\end{acknowledgments}

%\bibliographystyle{apsrev4-1}
%\bibliography{paperbib}

\begin{thebibliography}{43}%
\makeatletter
\providecommand \@ifxundefined [1]{%
 \@ifx{#1\undefined}
}%
\providecommand \@ifnum [1]{%
 \ifnum #1\expandafter \@firstoftwo
 \else \expandafter \@secondoftwo
 \fi
}%
\providecommand \@ifx [1]{%
 \ifx #1\expandafter \@firstoftwo
 \else \expandafter \@secondoftwo
 \fi
}%
\providecommand \natexlab [1]{#1}%
\providecommand \enquote  [1]{``#1''}%
\providecommand \bibnamefont  [1]{#1}%
\providecommand \bibfnamefont [1]{#1}%
\providecommand \citenamefont [1]{#1}%
\providecommand \href@noop [0]{\@secondoftwo}%
\providecommand \href [0]{\begingroup \@sanitize@url \@href}%
\providecommand \@href[1]{\@@startlink{#1}\@@href}%
\providecommand \@@href[1]{\endgroup#1\@@endlink}%
\providecommand \@sanitize@url [0]{\catcode `\\12\catcode `\$12\catcode
  `\&12\catcode `\#12\catcode `\^12\catcode `\_12\catcode `\%12\relax}%
\providecommand \@@startlink[1]{}%
\providecommand \@@endlink[0]{}%
\providecommand \url  [0]{\begingroup\@sanitize@url \@url }%
\providecommand \@url [1]{\endgroup\@href {#1}{\urlprefix }}%
\providecommand \urlprefix  [0]{URL }%
\providecommand \Eprint [0]{\href }%
\providecommand \doibase [0]{http://dx.doi.org/}%
\providecommand \selectlanguage [0]{\@gobble}%
\providecommand \bibinfo  [0]{\@secondoftwo}%
\providecommand \bibfield  [0]{\@secondoftwo}%
\providecommand \translation [1]{[#1]}%
\providecommand \BibitemOpen [0]{}%
\providecommand \bibitemStop [0]{}%
\providecommand \bibitemNoStop [0]{.\EOS\space}%
\providecommand \EOS [0]{\spacefactor3000\relax}%
\providecommand \BibitemShut  [1]{\csname bibitem#1\endcsname}%
\let\auto@bib@innerbib\@empty
%</preamble>
\bibitem [{\citenamefont {Drozdov}\ \emph {et~al.}(2015)\citenamefont
  {Drozdov}, \citenamefont {Eremets}, \citenamefont {Troyan}, \citenamefont
  {Ksenofontov},\ and\ \citenamefont {Shylin}}]{Drozdov:2015}%
  \BibitemOpen
  \bibfield  {author} {\bibinfo {author} {\bibfnamefont {AP}~\bibnamefont
  {Drozdov}}, \bibinfo {author} {\bibfnamefont {MI}~\bibnamefont {Eremets}},
  \bibinfo {author} {\bibfnamefont {IA}~\bibnamefont {Troyan}}, \bibinfo
  {author} {\bibfnamefont {V}~\bibnamefont {Ksenofontov}}, \ and\ \bibinfo
  {author} {\bibfnamefont {SI}~\bibnamefont {Shylin}},\ }\bibfield  {title}
  {\enquote {\bibinfo {title} {Conventional superconductivity at 203 kelvin at
  high pressures in the sulfur hydride system},}\ }\href@noop {} {\bibfield
  {journal} {\bibinfo  {journal} {Nature (London)}\ }\textbf {\bibinfo {volume}
  {525}},\ \bibinfo {pages} {73--76} (\bibinfo {year} {2015})}\BibitemShut
  {NoStop}%
\bibitem [{\citenamefont {Drozdov}\ \emph {et~al.}(2014)\citenamefont
  {Drozdov}, \citenamefont {Eremets},\ and\ \citenamefont
  {Troyan}}]{Drozdov:2014}%
  \BibitemOpen
  \bibfield  {author} {\bibinfo {author} {\bibfnamefont {A.~P.}\ \bibnamefont
  {Drozdov}}, \bibinfo {author} {\bibfnamefont {M.~I.}\ \bibnamefont
  {Eremets}}, \ and\ \bibinfo {author} {\bibfnamefont {I.~A.}\ \bibnamefont
  {Troyan}},\ }\href@noop {} {\enquote {\bibinfo {title} {Conventional
  superconductivity at 190 {K} at high pressures},}\ } (\bibinfo {year}
  {2014}),\ \Eprint {http://arxiv.org/abs/arXiv:1412.0460} {arXiv:1412.0460}
  \BibitemShut {NoStop}%
\bibitem [{\citenamefont {Schilling}\ \emph {et~al.}(1993)\citenamefont
  {Schilling}, \citenamefont {Cantoni}, \citenamefont {Guo},\ and\
  \citenamefont {Ott}}]{Schilling:1993}%
  \BibitemOpen
  \bibfield  {author} {\bibinfo {author} {\bibfnamefont {A.}~\bibnamefont
  {Schilling}}, \bibinfo {author} {\bibfnamefont {M.}~\bibnamefont {Cantoni}},
  \bibinfo {author} {\bibfnamefont {J.~D.}\ \bibnamefont {Guo}}, \ and\
  \bibinfo {author} {\bibfnamefont {H.~R.}\ \bibnamefont {Ott}},\ }\bibfield
  {title} {\enquote {\bibinfo {title} {{Superconductivity above 130 K in the
  Hg-Ba-Ca-Cu-O system}},}\ }\href@noop {} {\bibfield  {journal} {\bibinfo
  {journal} {Nature (London)}\ }\textbf {\bibinfo {volume} {363}},\ \bibinfo
  {pages} {56--58} (\bibinfo {year} {1993})}\BibitemShut {NoStop}%
\bibitem [{\citenamefont {Gao}\ \emph {et~al.}(1994)\citenamefont {Gao},
  \citenamefont {Xue}, \citenamefont {Chen}, \citenamefont {Xiong},
  \citenamefont {Meng}, \citenamefont {Ramirez}, \citenamefont {Chu},
  \citenamefont {Eggert},\ and\ \citenamefont {Mao}}]{Gao:1994}%
  \BibitemOpen
  \bibfield  {author} {\bibinfo {author} {\bibfnamefont {L.}~\bibnamefont
  {Gao}}, \bibinfo {author} {\bibfnamefont {Y.~Y.}\ \bibnamefont {Xue}},
  \bibinfo {author} {\bibfnamefont {F.}~\bibnamefont {Chen}}, \bibinfo {author}
  {\bibfnamefont {Q.}~\bibnamefont {Xiong}}, \bibinfo {author} {\bibfnamefont
  {R.~L.}\ \bibnamefont {Meng}}, \bibinfo {author} {\bibfnamefont
  {D.}~\bibnamefont {Ramirez}}, \bibinfo {author} {\bibfnamefont {C.~W.}\
  \bibnamefont {Chu}}, \bibinfo {author} {\bibfnamefont {J.~H.}\ \bibnamefont
  {Eggert}}, \ and\ \bibinfo {author} {\bibfnamefont {H.~K.}\ \bibnamefont
  {Mao}},\ }\bibfield  {title} {\enquote {\bibinfo {title} {{Superconductivity
  up to 164 K in
  ${\mathrm{HgBa}}_{2}$${\mathrm{Ca}}_{\mathit{m}\mathrm{\ensuremath{-}}1}$${\mathrm{Cu}}_{\mathit{m}}$${\mathrm{O}}_{2\mathit{m}+2+\mathrm{\ensuremath{\delta}}}$
  (m=1, 2, and 3) under quasihydrostatic pressures}},}\ }\href {\doibase
  10.1103/PhysRevB.50.4260} {\bibfield  {journal} {\bibinfo  {journal} {Phys.
  Rev. B}\ }\textbf {\bibinfo {volume} {50}},\ \bibinfo {pages} {4260--4263}
  (\bibinfo {year} {1994})}\BibitemShut {NoStop}%
\bibitem [{\citenamefont {O'Donovan}\ and\ \citenamefont
  {Carbotte}(1995{\natexlab{a}})}]{ODonovan:1995a}%
  \BibitemOpen
  \bibfield  {author} {\bibinfo {author} {\bibfnamefont {C.}~\bibnamefont
  {O'Donovan}}\ and\ \bibinfo {author} {\bibfnamefont {J.~P.}\ \bibnamefont
  {Carbotte}},\ }\bibfield  {title} {\enquote {\bibinfo {title} {In-plane
  penetration-depth anisotropy in a d-wave model},}\ }\href {\doibase
  10.1103/PhysRevB.52.4568} {\bibfield  {journal} {\bibinfo  {journal} {Phys.
  Rev. B}\ }\textbf {\bibinfo {volume} {52}},\ \bibinfo {pages} {4568--4576}
  (\bibinfo {year} {1995}{\natexlab{a}})}\BibitemShut {NoStop}%
\bibitem [{\citenamefont {O'Donovan}\ and\ \citenamefont
  {Carbotte}(1995{\natexlab{b}})}]{ODonovan:1995b}%
  \BibitemOpen
  \bibfield  {author} {\bibinfo {author} {\bibfnamefont {C.}~\bibnamefont
  {O'Donovan}}\ and\ \bibinfo {author} {\bibfnamefont {J.~P.}\ \bibnamefont
  {Carbotte}},\ }\bibfield  {title} {\enquote {\bibinfo {title} {Mixed order
  parameter symmetry in the {BCS} model},}\ }\href@noop {} {\bibfield
  {journal} {\bibinfo  {journal} {Physica C}\ }\textbf {\bibinfo {volume}
  {252}},\ \bibinfo {pages} {87} (\bibinfo {year}
  {1995}{\natexlab{b}})}\BibitemShut {NoStop}%
\bibitem [{\citenamefont {Carbotte}\ \emph {et~al.}(2011)\citenamefont
  {Carbotte}, \citenamefont {Timusk},\ and\ \citenamefont
  {Hwang}}]{Carbotte:2011}%
  \BibitemOpen
  \bibfield  {author} {\bibinfo {author} {\bibfnamefont {J.~P.}\ \bibnamefont
  {Carbotte}}, \bibinfo {author} {\bibfnamefont {T.}~\bibnamefont {Timusk}}, \
  and\ \bibinfo {author} {\bibfnamefont {J.}~\bibnamefont {Hwang}},\ }\bibfield
   {title} {\enquote {\bibinfo {title} {Bosons in high-temperature
  superconductors: {An} experimental survey},}\ }\href@noop {} {\bibfield
  {journal} {\bibinfo  {journal} {Rep. Prog. Phys.}\ }\textbf {\bibinfo
  {volume} {74}},\ \bibinfo {pages} {066501} (\bibinfo {year}
  {2011})}\BibitemShut {NoStop}%
\bibitem [{\citenamefont {Leung}\ \emph {et~al.}(1976)\citenamefont {Leung},
  \citenamefont {Carbotte}, \citenamefont {Taylor},\ and\ \citenamefont
  {Leavens}}]{Leung:1976}%
  \BibitemOpen
  \bibfield  {author} {\bibinfo {author} {\bibfnamefont {H.~K.}\ \bibnamefont
  {Leung}}, \bibinfo {author} {\bibfnamefont {J.~P.}\ \bibnamefont {Carbotte}},
  \bibinfo {author} {\bibfnamefont {D.~W.}\ \bibnamefont {Taylor}}, \ and\
  \bibinfo {author} {\bibfnamefont {C.~R.}\ \bibnamefont {Leavens}},\
  }\bibfield  {title} {\enquote {\bibinfo {title} {{Multiple-plane wave
  calculations of the electron-phonon interaction in Al}},}\ }\href {\doibase
  10.1139/p76-187} {\bibfield  {journal} {\bibinfo  {journal} {Can. J. Phys.}\
  }\textbf {\bibinfo {volume} {54}},\ \bibinfo {pages} {1585--1599} (\bibinfo
  {year} {1976})}\BibitemShut {NoStop}%
\bibitem [{\citenamefont {Cohen}\ and\ \citenamefont
  {Anderson}(1972)}]{Cohen:1972}%
  \BibitemOpen
  \bibfield  {author} {\bibinfo {author} {\bibfnamefont {M.~L.}\ \bibnamefont
  {Cohen}}\ and\ \bibinfo {author} {\bibfnamefont {P.~W.}\ \bibnamefont
  {Anderson}},\ }\bibfield  {title} {\enquote {\bibinfo {title} {Comments on
  the maximum superconducting transition temperature},}\ }in\ \href@noop {}
  {\emph {\bibinfo {booktitle} {d- and f-band Metals}}},\ Vol.~\bibinfo
  {volume} {4},\ \bibinfo {editor} {edited by\ \bibinfo {editor} {\bibfnamefont
  {D.~H.}\ \bibnamefont {Douglass}}}\ (\bibinfo  {publisher} {AIP},\ \bibinfo
  {address} {New York},\ \bibinfo {year} {1972})\ pp.\ \bibinfo {pages}
  {17--27}\BibitemShut {NoStop}%
\bibitem [{\citenamefont {Nagamatsu}\ \emph {et~al.}(2001)\citenamefont
  {Nagamatsu}, \citenamefont {Nakagawa}, \citenamefont {Muranaka},
  \citenamefont {Zenitani},\ and\ \citenamefont {Akimitsu}}]{Nagamatsu:2001}%
  \BibitemOpen
  \bibfield  {author} {\bibinfo {author} {\bibfnamefont {J.}~\bibnamefont
  {Nagamatsu}}, \bibinfo {author} {\bibfnamefont {N.}~\bibnamefont {Nakagawa}},
  \bibinfo {author} {\bibfnamefont {T.}~\bibnamefont {Muranaka}}, \bibinfo
  {author} {\bibfnamefont {Y.}~\bibnamefont {Zenitani}}, \ and\ \bibinfo
  {author} {\bibfnamefont {J.}~\bibnamefont {Akimitsu}},\ }\bibfield  {title}
  {\enquote {\bibinfo {title} {{Superconductivity at 39 K in magnesium
  diboride}},}\ }\href@noop {} {\bibfield  {journal} {\bibinfo  {journal}
  {Nature (London)}\ }\textbf {\bibinfo {volume} {410}},\ \bibinfo {pages} {63}
  (\bibinfo {year} {2001})}\BibitemShut {NoStop}%
\bibitem [{\citenamefont {Kortus}\ \emph {et~al.}(2001)\citenamefont {Kortus},
  \citenamefont {Mazin}, \citenamefont {Belashchenko}, \citenamefont
  {Antropov},\ and\ \citenamefont {Boyer}}]{Kortus:2001}%
  \BibitemOpen
  \bibfield  {author} {\bibinfo {author} {\bibfnamefont {J.}~\bibnamefont
  {Kortus}}, \bibinfo {author} {\bibfnamefont {I.~I.}\ \bibnamefont {Mazin}},
  \bibinfo {author} {\bibfnamefont {K.~D.}\ \bibnamefont {Belashchenko}},
  \bibinfo {author} {\bibfnamefont {V.~P.}\ \bibnamefont {Antropov}}, \ and\
  \bibinfo {author} {\bibfnamefont {L.~L.}\ \bibnamefont {Boyer}},\ }\bibfield
  {title} {\enquote {\bibinfo {title} {{Superconductivity of Metallic Boron in
  ${\mathrm{MgB}}_{2}$}},}\ }\href {\doibase 10.1103/PhysRevLett.86.4656}
  {\bibfield  {journal} {\bibinfo  {journal} {Phys. Rev. Lett.}\ }\textbf
  {\bibinfo {volume} {86}},\ \bibinfo {pages} {4656--4659} (\bibinfo {year}
  {2001})}\BibitemShut {NoStop}%
\bibitem [{\citenamefont {An}\ and\ \citenamefont {Pickett}(2001)}]{An:2001}%
  \BibitemOpen
  \bibfield  {author} {\bibinfo {author} {\bibfnamefont {J.~M.}\ \bibnamefont
  {An}}\ and\ \bibinfo {author} {\bibfnamefont {W.~E.}\ \bibnamefont
  {Pickett}},\ }\bibfield  {title} {\enquote {\bibinfo {title}
  {{Superconductivity of ${\mathrm{MgB}}_{2}$: Covalent Bonds Driven
  Metallic}},}\ }\href {\doibase 10.1103/PhysRevLett.86.4366} {\bibfield
  {journal} {\bibinfo  {journal} {Phys. Rev. Lett.}\ }\textbf {\bibinfo
  {volume} {86}},\ \bibinfo {pages} {4366--4369} (\bibinfo {year}
  {2001})}\BibitemShut {NoStop}%
\bibitem [{\citenamefont {Liu}\ \emph {et~al.}(2001)\citenamefont {Liu},
  \citenamefont {Mazin},\ and\ \citenamefont {Kortus}}]{Liu:2001}%
  \BibitemOpen
  \bibfield  {author} {\bibinfo {author} {\bibfnamefont {Amy~Y.}\ \bibnamefont
  {Liu}}, \bibinfo {author} {\bibfnamefont {I.~I.}\ \bibnamefont {Mazin}}, \
  and\ \bibinfo {author} {\bibfnamefont {Jens}\ \bibnamefont {Kortus}},\
  }\bibfield  {title} {\enquote {\bibinfo {title} {{Beyond Eliashberg
  Superconductivity in ${\mathrm{MgB}}_{2}$: Anharmonicity, Two-Phonon
  Scattering, and Multiple Gaps}},}\ }\href {\doibase
  10.1103/PhysRevLett.87.087005} {\bibfield  {journal} {\bibinfo  {journal}
  {Phys. Rev. Lett.}\ }\textbf {\bibinfo {volume} {87}},\ \bibinfo {pages}
  {087005} (\bibinfo {year} {2001})}\BibitemShut {NoStop}%
\bibitem [{\citenamefont {Kong}\ \emph {et~al.}(2001)\citenamefont {Kong},
  \citenamefont {Dolgov}, \citenamefont {Jepsen},\ and\ \citenamefont
  {Andersen}}]{Kong:2001}%
  \BibitemOpen
  \bibfield  {author} {\bibinfo {author} {\bibfnamefont {Y.}~\bibnamefont
  {Kong}}, \bibinfo {author} {\bibfnamefont {O.~V.}\ \bibnamefont {Dolgov}},
  \bibinfo {author} {\bibfnamefont {O.}~\bibnamefont {Jepsen}}, \ and\ \bibinfo
  {author} {\bibfnamefont {O.~K.}\ \bibnamefont {Andersen}},\ }\bibfield
  {title} {\enquote {\bibinfo {title} {{Electron-phonon interaction in the
  normal and superconducting states of ${\mathrm{MgB}}_{2}$}},}\ }\href
  {\doibase 10.1103/PhysRevB.64.020501} {\bibfield  {journal} {\bibinfo
  {journal} {Phys. Rev. B}\ }\textbf {\bibinfo {volume} {64}},\ \bibinfo
  {pages} {020501(R)} (\bibinfo {year} {2001})}\BibitemShut {NoStop}%
\bibitem [{\citenamefont {Choi}\ \emph {et~al.}(2002)\citenamefont {Choi},
  \citenamefont {Roundy}, \citenamefont {Sun}, \citenamefont {Cohen},\ and\
  \citenamefont {Louie}}]{Choi:2002}%
  \BibitemOpen
  \bibfield  {author} {\bibinfo {author} {\bibfnamefont {H.~J.}\ \bibnamefont
  {Choi}}, \bibinfo {author} {\bibfnamefont {D.}~\bibnamefont {Roundy}},
  \bibinfo {author} {\bibfnamefont {H.}~\bibnamefont {Sun}}, \bibinfo {author}
  {\bibfnamefont {M.~L.}\ \bibnamefont {Cohen}}, \ and\ \bibinfo {author}
  {\bibfnamefont {S.~G.}\ \bibnamefont {Louie}},\ }\bibfield  {title} {\enquote
  {\bibinfo {title} {{The origin of the anomalous superconducting properties of
  MgB$_2$}},}\ }\href@noop {} {\bibfield  {journal} {\bibinfo  {journal}
  {Nature (Nature)}\ }\textbf {\bibinfo {volume} {418}},\ \bibinfo {pages}
  {758} (\bibinfo {year} {2002})}\BibitemShut {NoStop}%
\bibitem [{\citenamefont {Nicol}\ and\ \citenamefont
  {Carbotte}(2005)}]{Nicol:2005}%
  \BibitemOpen
  \bibfield  {author} {\bibinfo {author} {\bibfnamefont {E.~J.}\ \bibnamefont
  {Nicol}}\ and\ \bibinfo {author} {\bibfnamefont {J.~P.}\ \bibnamefont
  {Carbotte}},\ }\bibfield  {title} {\enquote {\bibinfo {title} {Properties of
  the superconducting state in a two-band model},}\ }\href {\doibase
  10.1103/PhysRevB.71.054501} {\bibfield  {journal} {\bibinfo  {journal} {Phys.
  Rev. B}\ }\textbf {\bibinfo {volume} {71}},\ \bibinfo {pages} {054501}
  (\bibinfo {year} {2005})}\BibitemShut {NoStop}%
\bibitem [{\citenamefont {Ashcroft}(2004)}]{Ashcroft:2004}%
  \BibitemOpen
  \bibfield  {author} {\bibinfo {author} {\bibfnamefont {N.~W.}\ \bibnamefont
  {Ashcroft}},\ }\bibfield  {title} {\enquote {\bibinfo {title} {{Hydrogen
  Dominant Metallic Alloys: High Temperature Superconductors?}}}\ }\href
  {\doibase 10.1103/PhysRevLett.92.187002} {\bibfield  {journal} {\bibinfo
  {journal} {Phys. Rev. Lett.}\ }\textbf {\bibinfo {volume} {92}},\ \bibinfo
  {pages} {187002} (\bibinfo {year} {2004})}\BibitemShut {NoStop}%
\bibitem [{\citenamefont {Li}\ \emph {et~al.}(2014)\citenamefont {Li},
  \citenamefont {Hao}, \citenamefont {Lui}, \citenamefont {Li},\ and\
  \citenamefont {Ma}}]{Li:2014}%
  \BibitemOpen
  \bibfield  {author} {\bibinfo {author} {\bibfnamefont {Y.}~\bibnamefont
  {Li}}, \bibinfo {author} {\bibfnamefont {J.}~\bibnamefont {Hao}}, \bibinfo
  {author} {\bibfnamefont {H.}~\bibnamefont {Lui}}, \bibinfo {author}
  {\bibfnamefont {Y.}~\bibnamefont {Li}}, \ and\ \bibinfo {author}
  {\bibfnamefont {Y.}~\bibnamefont {Ma}},\ }\bibfield  {title} {\enquote
  {\bibinfo {title} {{The metallization and superconductivity of dense hydrogen
  sulfide}},}\ }\href@noop {} {\bibfield  {journal} {\bibinfo  {journal} {J.
  Chem. Phys.}\ }\textbf {\bibinfo {volume} {140}},\ \bibinfo {pages} {174712}
  (\bibinfo {year} {2014})}\BibitemShut {NoStop}%
\bibitem [{\citenamefont {Duan}\ \emph {et~al.}(2014)\citenamefont {Duan},
  \citenamefont {Liu}, \citenamefont {Tian}, \citenamefont {Li}, \citenamefont
  {Huang}, \citenamefont {Zhao}, \citenamefont {Yu}, \citenamefont {Liu},
  \citenamefont {Tian},\ and\ \citenamefont {Cui}}]{Duan:2014}%
  \BibitemOpen
  \bibfield  {author} {\bibinfo {author} {\bibfnamefont {Defang}\ \bibnamefont
  {Duan}}, \bibinfo {author} {\bibfnamefont {Yunxian}\ \bibnamefont {Liu}},
  \bibinfo {author} {\bibfnamefont {Fubo}\ \bibnamefont {Tian}}, \bibinfo
  {author} {\bibfnamefont {Da}~\bibnamefont {Li}}, \bibinfo {author}
  {\bibfnamefont {Xiaoli}\ \bibnamefont {Huang}}, \bibinfo {author}
  {\bibfnamefont {Zhonglong}\ \bibnamefont {Zhao}}, \bibinfo {author}
  {\bibfnamefont {Hongyu}\ \bibnamefont {Yu}}, \bibinfo {author} {\bibfnamefont
  {Bingbing}\ \bibnamefont {Liu}}, \bibinfo {author} {\bibfnamefont {Wenjing}\
  \bibnamefont {Tian}}, \ and\ \bibinfo {author} {\bibfnamefont {Tian}\
  \bibnamefont {Cui}},\ }\bibfield  {title} {\enquote {\bibinfo {title}
  {Pressure-induced metallization of dense {(H$_2$S)$_2$H$_2$} with
  {high-T$_c$} superconductivity},}\ }\href {\doibase 10.1038/srep06968}
  {\bibfield  {journal} {\bibinfo  {journal} {Sci. Rep.}\ }\textbf {\bibinfo
  {volume} {4}},\ \bibinfo {pages} {6968} (\bibinfo {year} {2014})}\BibitemShut
  {NoStop}%
\bibitem [{\citenamefont {Errea}\ \emph {et~al.}(2015)\citenamefont {Errea},
  \citenamefont {Calandra}, \citenamefont {Pickard}, \citenamefont {Nelson},
  \citenamefont {Needs}, \citenamefont {Li}, \citenamefont {Liu}, \citenamefont
  {Zhang}, \citenamefont {Ma},\ and\ \citenamefont {Mauri}}]{Errea:2015}%
  \BibitemOpen
  \bibfield  {author} {\bibinfo {author} {\bibfnamefont {Ion}\ \bibnamefont
  {Errea}}, \bibinfo {author} {\bibfnamefont {Matteo}\ \bibnamefont
  {Calandra}}, \bibinfo {author} {\bibfnamefont {Chris~J}\ \bibnamefont
  {Pickard}}, \bibinfo {author} {\bibfnamefont {Joseph~R}\ \bibnamefont
  {Nelson}}, \bibinfo {author} {\bibfnamefont {Richard~J}\ \bibnamefont
  {Needs}}, \bibinfo {author} {\bibfnamefont {Yinwei}\ \bibnamefont {Li}},
  \bibinfo {author} {\bibfnamefont {Hanyu}\ \bibnamefont {Liu}}, \bibinfo
  {author} {\bibfnamefont {Yunwei}\ \bibnamefont {Zhang}}, \bibinfo {author}
  {\bibfnamefont {Yanming}\ \bibnamefont {Ma}}, \ and\ \bibinfo {author}
  {\bibfnamefont {Francesco}\ \bibnamefont {Mauri}},\ }\bibfield  {title}
  {\enquote {\bibinfo {title} {High-pressure hydrogen sulfide from first
  principles: a strongly anharmonic phonon-mediated superconductor},}\
  }\href@noop {} {\bibfield  {journal} {\bibinfo  {journal} {Phys. Rev. Lett.}\
  }\textbf {\bibinfo {volume} {114}},\ \bibinfo {pages} {157004} (\bibinfo
  {year} {2015})}\BibitemShut {NoStop}%
\bibitem [{\citenamefont {Bernstein}\ \emph {et~al.}(2015)\citenamefont
  {Bernstein}, \citenamefont {Hellberg}, \citenamefont {Johannes},
  \citenamefont {Mazin},\ and\ \citenamefont {Mehl}}]{Bernstein:2015}%
  \BibitemOpen
  \bibfield  {author} {\bibinfo {author} {\bibfnamefont {N.}~\bibnamefont
  {Bernstein}}, \bibinfo {author} {\bibfnamefont {C.~Stephen}\ \bibnamefont
  {Hellberg}}, \bibinfo {author} {\bibfnamefont {M.~D.}\ \bibnamefont
  {Johannes}}, \bibinfo {author} {\bibfnamefont {I.~I.}\ \bibnamefont {Mazin}},
  \ and\ \bibinfo {author} {\bibfnamefont {M.~J.}\ \bibnamefont {Mehl}},\
  }\bibfield  {title} {\enquote {\bibinfo {title} {What superconducts in sulfur
  hydrides under pressure and why},}\ }\href {\doibase
  10.1103/PhysRevB.91.060511} {\bibfield  {journal} {\bibinfo  {journal} {Phys.
  Rev. B}\ }\textbf {\bibinfo {volume} {91}},\ \bibinfo {pages} {060511(R)}
  (\bibinfo {year} {2015})}\BibitemShut {NoStop}%
\bibitem [{\citenamefont {Papaconstantopoulos}\ \emph
  {et~al.}(2015)\citenamefont {Papaconstantopoulos}, \citenamefont {Klein},
  \citenamefont {Mehl},\ and\ \citenamefont
  {Pickett}}]{Papaconstantopoulos:2015}%
  \BibitemOpen
  \bibfield  {author} {\bibinfo {author} {\bibfnamefont {DA}~\bibnamefont
  {Papaconstantopoulos}}, \bibinfo {author} {\bibfnamefont {BM}~\bibnamefont
  {Klein}}, \bibinfo {author} {\bibfnamefont {MJ}~\bibnamefont {Mehl}}, \ and\
  \bibinfo {author} {\bibfnamefont {WE}~\bibnamefont {Pickett}},\ }\bibfield
  {title} {\enquote {\bibinfo {title} {{Cubic H$_3$S around 200 GPa: An atomic
  hydrogen superconductor stabilized by sulfur}},}\ }\href@noop {} {\bibfield
  {journal} {\bibinfo  {journal} {Phys. Rev. B}\ }\textbf {\bibinfo {volume}
  {91}},\ \bibinfo {pages} {184511} (\bibinfo {year} {2015})}\BibitemShut
  {NoStop}%
\bibitem [{\citenamefont {Flores-Livas}\ \emph {et~al.}(2016)\citenamefont
  {Flores-Livas}, \citenamefont {Sanna},\ and\ \citenamefont
  {Gross}}]{Flores-Livas:2016}%
  \BibitemOpen
  \bibfield  {author} {\bibinfo {author} {\bibfnamefont {Jos{\'e}~A}\
  \bibnamefont {Flores-Livas}}, \bibinfo {author} {\bibfnamefont {Antonio}\
  \bibnamefont {Sanna}}, \ and\ \bibinfo {author} {\bibfnamefont {EKU}\
  \bibnamefont {Gross}},\ }\bibfield  {title} {\enquote {\bibinfo {title} {High
  temperature superconductivity in sulfur and selenium hydrides at high
  pressure},}\ }\href@noop {} {\bibfield  {journal} {\bibinfo  {journal} {The
  European Physical Journal B}\ }\textbf {\bibinfo {volume} {89}},\ \bibinfo
  {pages} {1--6} (\bibinfo {year} {2016})}\BibitemShut {NoStop}%
\bibitem [{\citenamefont {Nicol}\ and\ \citenamefont
  {Carbotte}(2015)}]{Nicol:2015}%
  \BibitemOpen
  \bibfield  {author} {\bibinfo {author} {\bibfnamefont {EJ}~\bibnamefont
  {Nicol}}\ and\ \bibinfo {author} {\bibfnamefont {JP}~\bibnamefont
  {Carbotte}},\ }\bibfield  {title} {\enquote {\bibinfo {title} {Comparison of
  pressurized sulfur hydride with conventional superconductors},}\ }\href@noop
  {} {\bibfield  {journal} {\bibinfo  {journal} {Phys. Rev. B}\ }\textbf
  {\bibinfo {volume} {91}},\ \bibinfo {pages} {220507(R)} (\bibinfo {year}
  {2015})}\BibitemShut {NoStop}%
\bibitem [{\citenamefont {Errea}\ \emph {et~al.}(2016)\citenamefont {Errea},
  \citenamefont {Calandra}, \citenamefont {Pickard}, \citenamefont {Nelson},
  \citenamefont {Needs}, \citenamefont {Li}, \citenamefont {Liu}, \citenamefont
  {Zhang}, \citenamefont {Ma},\ and\ \citenamefont {Mauri}}]{Errea:2016}%
  \BibitemOpen
  \bibfield  {author} {\bibinfo {author} {\bibfnamefont {Ion}\ \bibnamefont
  {Errea}}, \bibinfo {author} {\bibfnamefont {Matteo}\ \bibnamefont
  {Calandra}}, \bibinfo {author} {\bibfnamefont {Chris~J}\ \bibnamefont
  {Pickard}}, \bibinfo {author} {\bibfnamefont {Joseph~R}\ \bibnamefont
  {Nelson}}, \bibinfo {author} {\bibfnamefont {Richard~J}\ \bibnamefont
  {Needs}}, \bibinfo {author} {\bibfnamefont {Yinwei}\ \bibnamefont {Li}},
  \bibinfo {author} {\bibfnamefont {Hanyu}\ \bibnamefont {Liu}}, \bibinfo
  {author} {\bibfnamefont {Yunwei}\ \bibnamefont {Zhang}}, \bibinfo {author}
  {\bibfnamefont {Yanming}\ \bibnamefont {Ma}}, \ and\ \bibinfo {author}
  {\bibfnamefont {Francesco}\ \bibnamefont {Mauri}},\ }\bibfield  {title}
  {\enquote {\bibinfo {title} {Quantum hydrogen-bond symmetrization in the
  superconducting hydrogen sulfide system},}\ }\href@noop {} {\bibfield
  {journal} {\bibinfo  {journal} {Nature (London)}\ }\textbf {\bibinfo {volume}
  {532}},\ \bibinfo {pages} {81--84} (\bibinfo {year} {2016})}\BibitemShut
  {NoStop}%
\bibitem [{\citenamefont {Einaga}\ \emph {et~al.}(2016)\citenamefont {Einaga},
  \citenamefont {Sakata}, \citenamefont {Ishikawa}, \citenamefont {Shimizu},
  \citenamefont {Eremets}, \citenamefont {Drozdov}, \citenamefont {Troyan},
  \citenamefont {Hirao},\ and\ \citenamefont {Ohishi}}]{Einaga:2016}%
  \BibitemOpen
  \bibfield  {author} {\bibinfo {author} {\bibfnamefont {Mari}\ \bibnamefont
  {Einaga}}, \bibinfo {author} {\bibfnamefont {Masafumi}\ \bibnamefont
  {Sakata}}, \bibinfo {author} {\bibfnamefont {Takahiro}\ \bibnamefont
  {Ishikawa}}, \bibinfo {author} {\bibfnamefont {Katsuya}\ \bibnamefont
  {Shimizu}}, \bibinfo {author} {\bibfnamefont {Mikhail~I}\ \bibnamefont
  {Eremets}}, \bibinfo {author} {\bibfnamefont {Alexander~P}\ \bibnamefont
  {Drozdov}}, \bibinfo {author} {\bibfnamefont {Ivan~A}\ \bibnamefont
  {Troyan}}, \bibinfo {author} {\bibfnamefont {Naohisa}\ \bibnamefont {Hirao}},
  \ and\ \bibinfo {author} {\bibfnamefont {Yasuo}\ \bibnamefont {Ohishi}},\
  }\bibfield  {title} {\enquote {\bibinfo {title} {Crystal structure of the
  superconducting phase of sulfur hydride},}\ }\href@noop {} {\bibfield
  {journal} {\bibinfo  {journal} {Nat. Phys.}\ }\textbf {\bibinfo {volume}
  {12}},\ \bibinfo {pages} {835--838} (\bibinfo {year} {2016})}\BibitemShut
  {NoStop}%
\bibitem [{\citenamefont {McMillan}\ and\ \citenamefont
  {Rowell}(1965)}]{McMillan:1965}%
  \BibitemOpen
  \bibfield  {author} {\bibinfo {author} {\bibfnamefont {W.~L.}\ \bibnamefont
  {McMillan}}\ and\ \bibinfo {author} {\bibfnamefont {J.~M.}\ \bibnamefont
  {Rowell}},\ }\bibfield  {title} {\enquote {\bibinfo {title} {Lead phonon
  spectrum calculated from superconducting density of states},}\ }\href
  {\doibase 10.1103/PhysRevLett.14.108} {\bibfield  {journal} {\bibinfo
  {journal} {Phys. Rev. Lett.}\ }\textbf {\bibinfo {volume} {14}},\ \bibinfo
  {pages} {108--112} (\bibinfo {year} {1965})}\BibitemShut {NoStop}%
\bibitem [{\citenamefont {Carbotte}(1990)}]{Carbotte:1990}%
  \BibitemOpen
  \bibfield  {author} {\bibinfo {author} {\bibfnamefont {J.~P.}\ \bibnamefont
  {Carbotte}},\ }\bibfield  {title} {\enquote {\bibinfo {title} {Properties of
  boson-exchange superconductors},}\ }\href {\doibase
  10.1103/RevModPhys.62.1027} {\bibfield  {journal} {\bibinfo  {journal} {Rev.
  Mod. Phys.}\ }\textbf {\bibinfo {volume} {62}},\ \bibinfo {pages}
  {1027--1157} (\bibinfo {year} {1990})}\BibitemShut {NoStop}%
\bibitem [{\citenamefont {Joyce}\ and\ \citenamefont
  {Richards}(1970)}]{Joyce:1970}%
  \BibitemOpen
  \bibfield  {author} {\bibinfo {author} {\bibfnamefont {R.~R.}\ \bibnamefont
  {Joyce}}\ and\ \bibinfo {author} {\bibfnamefont {P.~L.}\ \bibnamefont
  {Richards}},\ }\bibfield  {title} {\enquote {\bibinfo {title} {Phonon
  contribution to the far-infrared absorptivity of superconducting and normal
  lead},}\ }\href {\doibase 10.1103/PhysRevLett.24.1007} {\bibfield  {journal}
  {\bibinfo  {journal} {Phys. Rev. Lett.}\ }\textbf {\bibinfo {volume} {24}},\
  \bibinfo {pages} {1007--1011} (\bibinfo {year} {1970})}\BibitemShut {NoStop}%
\bibitem [{\citenamefont {Farnworth}\ and\ \citenamefont
  {Timusk}(1976)}]{Farnworth:1976}%
  \BibitemOpen
  \bibfield  {author} {\bibinfo {author} {\bibfnamefont {B.}~\bibnamefont
  {Farnworth}}\ and\ \bibinfo {author} {\bibfnamefont {T.}~\bibnamefont
  {Timusk}},\ }\bibfield  {title} {\enquote {\bibinfo {title} {Phonon density
  of states of superconducting lead},}\ }\href {\doibase
  10.1103/PhysRevB.14.5119} {\bibfield  {journal} {\bibinfo  {journal} {Phys.
  Rev. B}\ }\textbf {\bibinfo {volume} {14}},\ \bibinfo {pages} {5119--5120}
  (\bibinfo {year} {1976})}\BibitemShut {NoStop}%
\bibitem [{\citenamefont {Basov}\ and\ \citenamefont
  {Timusk}(2005)}]{Basov:2005}%
  \BibitemOpen
  \bibfield  {author} {\bibinfo {author} {\bibfnamefont {D.~N.}\ \bibnamefont
  {Basov}}\ and\ \bibinfo {author} {\bibfnamefont {T.}~\bibnamefont {Timusk}},\
  }\bibfield  {title} {\enquote {\bibinfo {title} {Electrodynamics of
  high-${T}_{c}$ superconductors},}\ }\href {\doibase
  10.1103/RevModPhys.77.721} {\bibfield  {journal} {\bibinfo  {journal} {Rev.
  Mod. Phys.}\ }\textbf {\bibinfo {volume} {77}},\ \bibinfo {pages} {721--779}
  (\bibinfo {year} {2005})}\BibitemShut {NoStop}%
\bibitem [{\citenamefont {F.}\ \emph {et~al.}(1998)\citenamefont {F.},
  \citenamefont {Startseva},\ and\ \citenamefont {Carbotte}}]{Marsiglio:1998}%
  \BibitemOpen
  \bibfield  {author} {\bibinfo {author} {\bibfnamefont {Marsiglio}\
  \bibnamefont {F.}}, \bibinfo {author} {\bibfnamefont {T.}~\bibnamefont
  {Startseva}}, \ and\ \bibinfo {author} {\bibfnamefont {J.~P.}\ \bibnamefont
  {Carbotte}},\ }\bibfield  {title} {\enquote {\bibinfo {title} {{Inversion of
  K$_3$C$_{60}$ reflectance data}},}\ }\href@noop {} {\bibfield  {journal}
  {\bibinfo  {journal} {Phys. Lett. A}\ }\textbf {\bibinfo {volume} {245}},\
  \bibinfo {pages} {172} (\bibinfo {year} {1998})}\BibitemShut {NoStop}%
\bibitem [{\citenamefont {Mori}\ \emph {et~al.}(2008)\citenamefont {Mori},
  \citenamefont {Nicol}, \citenamefont {Shiizuka}, \citenamefont {Kuniyasu},
  \citenamefont {Nojima}, \citenamefont {Toyota},\ and\ \citenamefont
  {Carbotte}}]{Mori:2008}%
  \BibitemOpen
  \bibfield  {author} {\bibinfo {author} {\bibfnamefont {T.}~\bibnamefont
  {Mori}}, \bibinfo {author} {\bibfnamefont {E.~J.}\ \bibnamefont {Nicol}},
  \bibinfo {author} {\bibfnamefont {S.}~\bibnamefont {Shiizuka}}, \bibinfo
  {author} {\bibfnamefont {K.}~\bibnamefont {Kuniyasu}}, \bibinfo {author}
  {\bibfnamefont {T.}~\bibnamefont {Nojima}}, \bibinfo {author} {\bibfnamefont
  {N.}~\bibnamefont {Toyota}}, \ and\ \bibinfo {author} {\bibfnamefont {J.~P.}\
  \bibnamefont {Carbotte}},\ }\bibfield  {title} {\enquote {\bibinfo {title}
  {Optical self-energy of superconducting {Pb} in the terahertz region},}\
  }\href {\doibase 10.1103/PhysRevB.77.174515} {\bibfield  {journal} {\bibinfo
  {journal} {Phys. Rev. B}\ }\textbf {\bibinfo {volume} {77}},\ \bibinfo
  {pages} {174515} (\bibinfo {year} {2008})}\BibitemShut {NoStop}%
\bibitem [{\citenamefont {Hwang}\ and\ \citenamefont
  {Carbotte}(2014)}]{Hwang:2014}%
  \BibitemOpen
  \bibfield  {author} {\bibinfo {author} {\bibfnamefont {J.}~\bibnamefont
  {Hwang}}\ and\ \bibinfo {author} {\bibfnamefont {J.~P.}\ \bibnamefont
  {Carbotte}},\ }\bibfield  {title} {\enquote {\bibinfo {title} {{Deriving the
  electron-phonon spectral density of MgB$_2$ from optical data, using maximum
  entropy techniques}},}\ }\href@noop {} {\bibfield  {journal} {\bibinfo
  {journal} {J. Phys. Condensed Matter}\ }\textbf {\bibinfo {volume} {26}},\
  \bibinfo {pages} {165702} (\bibinfo {year} {2014})}\BibitemShut {NoStop}%
\bibitem [{\citenamefont {Capitani}\ \emph {et~al.}(2017)\citenamefont
  {Capitani}, \citenamefont {Langerome}, \citenamefont {Brubach}, \citenamefont
  {Roy}, \citenamefont {Drozdov}, \citenamefont {Eremets}, \citenamefont
  {Nicol}, \citenamefont {Carbotte},\ and\ \citenamefont
  {Timusk}}]{Capitani:2017}%
  \BibitemOpen
  \bibfield  {author} {\bibinfo {author} {\bibfnamefont {F}~\bibnamefont
  {Capitani}}, \bibinfo {author} {\bibfnamefont {B}~\bibnamefont {Langerome}},
  \bibinfo {author} {\bibfnamefont {J-B}\ \bibnamefont {Brubach}}, \bibinfo
  {author} {\bibfnamefont {P}~\bibnamefont {Roy}}, \bibinfo {author}
  {\bibfnamefont {A}~\bibnamefont {Drozdov}}, \bibinfo {author} {\bibfnamefont
  {MI}~\bibnamefont {Eremets}}, \bibinfo {author} {\bibfnamefont
  {EJ}~\bibnamefont {Nicol}}, \bibinfo {author} {\bibfnamefont
  {JP}~\bibnamefont {Carbotte}}, \ and\ \bibinfo {author} {\bibfnamefont
  {T}~\bibnamefont {Timusk}},\ }\bibfield  {title} {\enquote {\bibinfo {title}
  {Spectroscopic evidence of a new energy scale for superconductivity in
  {H$_3$S}},}\ }\href@noop {} {\bibfield  {journal} {\bibinfo  {journal} {Nat.
  Phys.}\ }\textbf {\bibinfo {volume} {13}},\ \bibinfo {pages} {859} (\bibinfo
  {year} {2017})}\BibitemShut {NoStop}%
\bibitem [{\citenamefont {Marsiglio}\ and\ \citenamefont
  {Carbotte}(2008)}]{Marsiglio:2008}%
  \BibitemOpen
  \bibfield  {author} {\bibinfo {author} {\bibfnamefont {F.}~\bibnamefont
  {Marsiglio}}\ and\ \bibinfo {author} {\bibfnamefont {J.~P.}\ \bibnamefont
  {Carbotte}},\ }\bibfield  {title} {\enquote {\bibinfo {title}
  {Electron-phonon superconductivity},}\ }in\ \href@noop {} {\emph {\bibinfo
  {booktitle} {Superconductivity in Conventional and Unconventional
  Superconductors}}},\ \bibinfo {editor} {edited by\ \bibinfo {editor}
  {\bibfnamefont {K.H.}\ \bibnamefont {Bennemann}}\ and\ \bibinfo {editor}
  {\bibfnamefont {J.B.}\ \bibnamefont {Ketterson}}}\ (\bibinfo  {publisher}
  {Springer},\ \bibinfo {address} {Berlin},\ \bibinfo {year} {2008})\ pp.\
  \bibinfo {pages} {73--162}\BibitemShut {NoStop}%
\bibitem [{\citenamefont {Liu}\ \emph {et~al.}(2017)\citenamefont {Liu},
  \citenamefont {Naumov}, \citenamefont {Hoffmann}, \citenamefont {Ashcroft},\
  and\ \citenamefont {Hemley}}]{Liu:2017}%
  \BibitemOpen
  \bibfield  {author} {\bibinfo {author} {\bibfnamefont {Hanyu}\ \bibnamefont
  {Liu}}, \bibinfo {author} {\bibfnamefont {Ivan~I}\ \bibnamefont {Naumov}},
  \bibinfo {author} {\bibfnamefont {Roald}\ \bibnamefont {Hoffmann}}, \bibinfo
  {author} {\bibfnamefont {NW}~\bibnamefont {Ashcroft}}, \ and\ \bibinfo
  {author} {\bibfnamefont {Russell~J}\ \bibnamefont {Hemley}},\ }\bibfield
  {title} {\enquote {\bibinfo {title} {Potential high-{T$_c$} superconducting
  lanthanum and yttrium hydrides at high pressure},}\ }\href@noop {} {\bibfield
   {journal} {\bibinfo  {journal} {Proceedings of the National Academy of
  Sciences}\ }\textbf {\bibinfo {volume} {114}},\ \bibinfo {pages} {6990--6995}
  (\bibinfo {year} {2017})}\BibitemShut {NoStop}%
\bibitem [{\citenamefont {Somayazulu}\ \emph {et~al.}(2019)\citenamefont
  {Somayazulu}, \citenamefont {Ahart}, \citenamefont {Mishra}, \citenamefont
  {Geballe}, \citenamefont {Baldini}, \citenamefont {Meng}, \citenamefont
  {Struzhkin},\ and\ \citenamefont {Hemley}}]{Somayazulu:2019}%
  \BibitemOpen
  \bibfield  {author} {\bibinfo {author} {\bibfnamefont {Maddury}\ \bibnamefont
  {Somayazulu}}, \bibinfo {author} {\bibfnamefont {Muhtar}\ \bibnamefont
  {Ahart}}, \bibinfo {author} {\bibfnamefont {Ajay~K.}\ \bibnamefont {Mishra}},
  \bibinfo {author} {\bibfnamefont {Zachary~M.}\ \bibnamefont {Geballe}},
  \bibinfo {author} {\bibfnamefont {Maria}\ \bibnamefont {Baldini}}, \bibinfo
  {author} {\bibfnamefont {Yue}\ \bibnamefont {Meng}}, \bibinfo {author}
  {\bibfnamefont {Viktor~V.}\ \bibnamefont {Struzhkin}}, \ and\ \bibinfo
  {author} {\bibfnamefont {Russell~J.}\ \bibnamefont {Hemley}},\ }\bibfield
  {title} {\enquote {\bibinfo {title} {{Evidence for Superconductivity above
  260 K in Lanthanum Superhydride at Megabar Pressures}},}\ }\href {\doibase
  10.1103/PhysRevLett.122.027001} {\bibfield  {journal} {\bibinfo  {journal}
  {Phys. Rev. Lett.}\ }\textbf {\bibinfo {volume} {122}},\ \bibinfo {pages}
  {027001} (\bibinfo {year} {2019})}\BibitemShut {NoStop}%
\bibitem [{\citenamefont {Drozdov}\ \emph {et~al.}(2018)\citenamefont
  {Drozdov}, \citenamefont {Minkov}, \citenamefont {Besedin}, \citenamefont
  {Kong}, \citenamefont {Kuzovnikov}, \citenamefont {Knyazev},\ and\
  \citenamefont {Eremets}}]{Drozdov:2018a}%
  \BibitemOpen
  \bibfield  {author} {\bibinfo {author} {\bibfnamefont {A.~P.}\ \bibnamefont
  {Drozdov}}, \bibinfo {author} {\bibfnamefont {V.~S.}\ \bibnamefont {Minkov}},
  \bibinfo {author} {\bibfnamefont {S.~P.}\ \bibnamefont {Besedin}}, \bibinfo
  {author} {\bibfnamefont {P.~P.}\ \bibnamefont {Kong}}, \bibinfo {author}
  {\bibfnamefont {M.~A.}\ \bibnamefont {Kuzovnikov}}, \bibinfo {author}
  {\bibfnamefont {D.~A.}\ \bibnamefont {Knyazev}}, \ and\ \bibinfo {author}
  {\bibfnamefont {M.~I.}\ \bibnamefont {Eremets}},\ }\href@noop {} {\enquote
  {\bibinfo {title} {{Superconductivity at 215 K in lanthanum hydride at high
  pressure}},}\ } (\bibinfo {year} {2018}),\ \Eprint
  {http://arxiv.org/abs/arXiv:1808.07039} {arXiv:1808.07039} \BibitemShut
  {NoStop}%
\bibitem [{\citenamefont {Drozdov}\ \emph {et~al.}(2019)\citenamefont
  {Drozdov}, \citenamefont {Kong}, \citenamefont {Minkov}, \citenamefont
  {Besedin}, \citenamefont {Kuzovnikov}, \citenamefont {Mozaffari},
  \citenamefont {Balicas}, \citenamefont {Balakirev}, \citenamefont {Graf},
  \citenamefont {Prakapenka}, \citenamefont {Greenberg}, \citenamefont
  {Knyazev}, \citenamefont {Tkacz},\ and\ \citenamefont
  {Eremets}}]{Drozdov:2019}%
  \BibitemOpen
  \bibfield  {author} {\bibinfo {author} {\bibfnamefont {A.~P.}\ \bibnamefont
  {Drozdov}}, \bibinfo {author} {\bibfnamefont {P.~P.}\ \bibnamefont {Kong}},
  \bibinfo {author} {\bibfnamefont {V.~S.}\ \bibnamefont {Minkov}}, \bibinfo
  {author} {\bibfnamefont {S.~P.}\ \bibnamefont {Besedin}}, \bibinfo {author}
  {\bibfnamefont {M.~A.}\ \bibnamefont {Kuzovnikov}}, \bibinfo {author}
  {\bibfnamefont {S.}~\bibnamefont {Mozaffari}}, \bibinfo {author}
  {\bibfnamefont {L.}~\bibnamefont {Balicas}}, \bibinfo {author} {\bibfnamefont
  {F.~F.}\ \bibnamefont {Balakirev}}, \bibinfo {author} {\bibfnamefont {D.~E.}\
  \bibnamefont {Graf}}, \bibinfo {author} {\bibfnamefont {V.~B.}\ \bibnamefont
  {Prakapenka}}, \bibinfo {author} {\bibfnamefont {E.}~\bibnamefont
  {Greenberg}}, \bibinfo {author} {\bibfnamefont {D.~A.}\ \bibnamefont
  {Knyazev}}, \bibinfo {author} {\bibfnamefont {M.}~\bibnamefont {Tkacz}}, \
  and\ \bibinfo {author} {\bibfnamefont {M.~I.}\ \bibnamefont {Eremets}},\
  }\bibfield  {title} {\enquote {\bibinfo {title} {{Superconductivity at 250 K
  in lanthanum hydride at high pressure}},}\ }\href@noop {} {\bibfield
  {journal} {\bibinfo  {journal} {Nature (London)}\ }\textbf {\bibinfo {volume}
  {569}},\ \bibinfo {pages} {528--531} (\bibinfo {year} {2019})}\BibitemShut
  {NoStop}%
\bibitem [{\citenamefont {Marsiglio}\ \emph {et~al.}(1988)\citenamefont
  {Marsiglio}, \citenamefont {Schossmann},\ and\ \citenamefont
  {Carbotte}}]{Marsiglio:1988}%
  \BibitemOpen
  \bibfield  {author} {\bibinfo {author} {\bibfnamefont {F.}~\bibnamefont
  {Marsiglio}}, \bibinfo {author} {\bibfnamefont {M.}~\bibnamefont
  {Schossmann}}, \ and\ \bibinfo {author} {\bibfnamefont {J.~P.}\ \bibnamefont
  {Carbotte}},\ }\bibfield  {title} {\enquote {\bibinfo {title} {Iterative
  analytic continuation of the electron self-energy to the real axis},}\ }\href
  {\doibase 10.1103/PhysRevB.37.4965} {\bibfield  {journal} {\bibinfo
  {journal} {Phys. Rev. B}\ }\textbf {\bibinfo {volume} {37}},\ \bibinfo
  {pages} {4965--4969} (\bibinfo {year} {1988})}\BibitemShut {NoStop}%
\bibitem [{\citenamefont {Akis}\ \emph {et~al.}(1991)\citenamefont {Akis},
  \citenamefont {Carbotte},\ and\ \citenamefont {Timusk}}]{Akis:1991}%
  \BibitemOpen
  \bibfield  {author} {\bibinfo {author} {\bibfnamefont {R.}~\bibnamefont
  {Akis}}, \bibinfo {author} {\bibfnamefont {J.~P.}\ \bibnamefont {Carbotte}},
  \ and\ \bibinfo {author} {\bibfnamefont {T.}~\bibnamefont {Timusk}},\
  }\bibfield  {title} {\enquote {\bibinfo {title} {Superconducting optical
  conductivity for arbitrary temperature and mean free path},}\ }\href
  {\doibase 10.1103/PhysRevB.43.12804} {\bibfield  {journal} {\bibinfo
  {journal} {Phys. Rev. B}\ }\textbf {\bibinfo {volume} {43}},\ \bibinfo
  {pages} {12804--12808} (\bibinfo {year} {1991})}\BibitemShut {NoStop}%
\bibitem [{\citenamefont {Carbotte}\ \emph {et~al.}(2018)\citenamefont
  {Carbotte}, \citenamefont {Nicol},\ and\ \citenamefont
  {Timusk}}]{Carbotte:2018}%
  \BibitemOpen
  \bibfield  {author} {\bibinfo {author} {\bibfnamefont {J.~P.}\ \bibnamefont
  {Carbotte}}, \bibinfo {author} {\bibfnamefont {E.~J.}\ \bibnamefont {Nicol}},
  \ and\ \bibinfo {author} {\bibfnamefont {T.}~\bibnamefont {Timusk}},\
  }\bibfield  {title} {\enquote {\bibinfo {title} {{Detecting Superconductivity
  in the High Pressure Hydrides and Metallic Hydrogen from Optical
  Properties}},}\ }\href {\doibase 10.1103/PhysRevLett.121.047002} {\bibfield
  {journal} {\bibinfo  {journal} {Phys. Rev. Lett.}\ }\textbf {\bibinfo
  {volume} {121}},\ \bibinfo {pages} {047002} (\bibinfo {year}
  {2018})}\BibitemShut {NoStop}%
\end{thebibliography}


\end{document}